\documentclass[conference]{IEEEtran}
\IEEEoverridecommandlockouts

\usepackage{cite}
\usepackage{amsmath,amssymb,amsfonts}
\usepackage{algorithmic}
\usepackage{graphicx}
\usepackage{textcomp}
\usepackage{xcolor}
\def\BibTeX{{\rm B\kern-.05em{\sc i\kern-.025em b}\kern-.08em
    T\kern-.1667em\lower.7ex\hbox{E}\kern-.125emX}}

\usepackage{microtype}
\usepackage{lipsum}
\usepackage{algorithm}
\usepackage{multirow}
\usepackage{caption}
\usepackage{subcaption}
\usepackage{todonotes}
\usepackage{url}

\usepackage{verbatim}
\usepackage[]{xcolor}

\usepackage{tabularx}
\newcommand{\datasetwebsite}{\url{https://github.com/sbu-dsl/prosody-analysis-of-audiobooks}}

\def\BibTeX{{\rm B\kern-.05em{\sc i\kern-.025em b}\kern-.08em
    T\kern-.1667em\lower.7ex\hbox{E}\kern-.125emX}}
\begin{document}

\title{Prosody Analysis of Audiobooks\\
}

\makeatletter
\newcommand{\linebreakand}{%
  \end{@IEEEauthorhalign}
  \hfill\mbox{}\par
  \mbox{}\hfill\begin{@IEEEauthorhalign}
}
\makeatother

\author{\IEEEauthorblockN{Charuta Pethe}
\IEEEauthorblockA{\textit{Department of Computer Science} \\
\textit{Stony Brook University}\\
Stony Brook, NY, USA \\
cpethe@cs.stonybrook.edu}
\and
\IEEEauthorblockN{Bach Pham}
\IEEEauthorblockA{\textit{Department of Computer Science} \\
\textit{Earlham College}\\
Richmond, IN, USA \\
bqpham24@earlham.edu}
\and
\IEEEauthorblockN{Felix D Childress}
\IEEEauthorblockA{\textit{Department of Computer Science} \\
\textit{Earlham College}\\
Richmond, IN, USA \\
fdchild22@earlham.edu}
\linebreakand
\IEEEauthorblockN{Yunting Yin}
\IEEEauthorblockA{\textit{Department of Computer Science} \\
\textit{Earlham College}\\
Richmond, IN, USA \\
yinyu@earlham.edu}
\and
\IEEEauthorblockN{Steven Skiena}
\IEEEauthorblockA{\textit{Department of Computer Science} \\
\textit{Stony Brook University}\\
Stony Brook, NY, USA \\
skiena@cs.stonybrook.edu}
}

\maketitle

\begin{abstract}
Recent advances in text-to-speech have made it possible to generate natural-sounding audio from text. However, audiobook narrations involve dramatic vocalizations and intonations by the reader, with greater reliance on emotions, dialogues, and descriptions in the narrative. 
Using our dataset of 93 aligned book-audiobook pairs, we present improved models to predict prosody (pitch, volume, and rate of speech) from narrative text using language modeling.
Our predicted prosody attributes correlate much better with human audiobook readings than results from a state-of-the-art commercial TTS system: our predicted pitch shows a higher correlation with human reading for 22 out of 24 books in the test set, while our predicted volume attribute proves more similar to human reading for 23 out of the 24 books.
Finally, we present a human evaluation study to quantify the extent that people prefer prosody-enhanced audiobook readings over default commercial text-to-speech systems.
\end{abstract}


\section{Introduction}
Audio books are actor-narrated recordings of written texts, typically full length novels.
Audio books have grown rapidly in popularity in recent years, with annual sales in the United States reaching \$1.3 billion dollars in 2020.\footnote{\url{https://www.statista.com/topics/3296/audiobooks/}}.
The National Library Service for the Blind has recorded tens of thousands of talking books for the sight impaired since 1933, read by an extensive network of volunteers.

In this paper, we analyze audio books from an NLP perspective, with two distinct objectives in mind.
First, we study how higher-order NLP analysis of novels (e.g. character identification, quote/dialog analysis, and narrative flow analysis) might be used to build better TTS systems. Second, we are interested as audio books as a source of {\em annotations} for narrative texts.
One of the challenges of NLP on book-length documents is the cost of annotation: reading a novel is a 10-15 hour commitment, making it cost-prohibitive to do human annotation for special-purpose tasks on a large corpora of books. We see human-recorded audio books as a potential solution here: each human recording implicitly contains information about the contents of the texts, which we can programmatically extract by analysing the audio. Our main contributions in this work include:

\begin{itemize}
    
    \item \textbf{Character-level analysis of audiobook reader behaviors} ---
    We quantify the extent to which the gender properties of characters are reflected in audiobook readings. Specifically, in 21 of 31 books where the two lead characters differ in gender, readers used lower pitch and higher volume to portray the male character. Further, readers generally use lower pitch in narrative regions rather than dialog, independent of gender.
    
    \item \textbf{Models for audiobook prosody prediction} --- We address the task of predicting prosody attributes given a narrative text, training a variety of models on audiobook readings to predict prosody attributes (pitch, volume, and rate). The dot plot of Figure \ref{fig:corr_human} presents the human-TTS correlation for our system using a LSTM architecture built upon MPNet embeddings, and a state-of-the-art commercial TTS system (Google Cloud Text-to-Speech) on a test set of the first chapters of 24 books. Our predicted pitch attribute shows a higher correlation with human reading for 22 out of the 24 books, while our predicted volume attribute proves more similar to human reading for 23 out of the 24 books.
    
\begin{figure}[t]
\centering
\begin{subfigure}{\linewidth}
  \centering
  \includegraphics[width=0.8\linewidth]{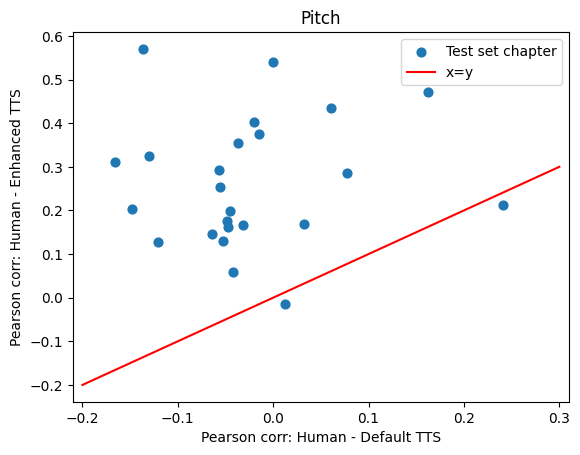}
  \caption{Pitch}
  \label{fig:pitch_corr_human}
\end{subfigure}%
\\
\begin{subfigure}{\linewidth}
  \centering
  \includegraphics[width=0.8\linewidth]{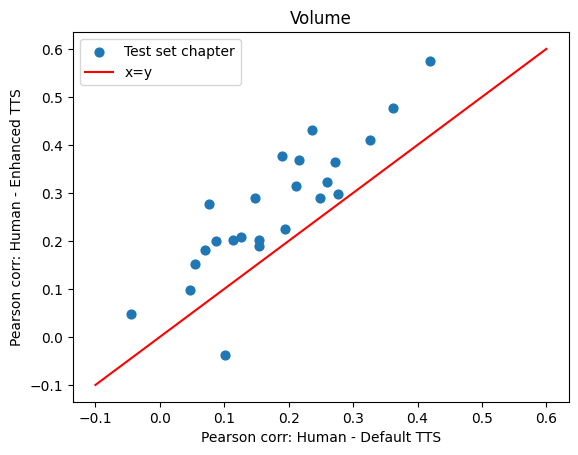}
  \caption{Volume}
  \label{fig:volume_corr_human}
\end{subfigure}
\caption{Enhancing TTS with our predicted prosody attributes show better correlations with human reading than unaided Google Cloud Text-to-Speech.}
\label{fig:corr_human}
\end{figure}

   \item \textbf{Human evaluation study for text-to-speech audiobook reading} ---
   We conduct a human evaluation study to understand whether humans prefer to hear our more expressive text-to-speech audio enhanced with Speech Synthesis Markup Language (SSML) over the default commercial Google Cloud Text-to-Speech, a strong baseline. Results are inconclusive, but a small majority (12 of 22) subjects preferred our SSML-enhanced readings.
\end{itemize}

Over the course of this study, we have developed a dataset of sentence-aligned books and audiobooks comprising 1806 chapters across 93 novels, with their corresponding human-read audiobooks and sentence-level alignment between the text and audio. This dataset has been made publicly available at \datasetwebsite.

This paper is organized as follows.
Section \ref{sec:related-work} presents related work associated with NLP for books and an overview of the Speech Synthesis Markup Language (SSML) we use in our experiments.
Section \ref{sec:audio-text-alignment} describes our methods for the critical phase of alignment between the textual representation of novels and associated audio book recordings.
Section \ref{sec:reader-behavior} presents our analysis of reader behavior, quantifying the degree to which voices change with a character's gender.
We develop prosody prediction models from text in Section \ref{sec:prosody-prediction},
culminating in a human preference study reported in Section \ref{sec:human-evaluation}.
Future directions for our work are discussed in Section \ref{sec:conclusions}.

\section{Related Work}
\label{sec:related-work}

\subsection{NLP for books:}
Recent work focused on performing NLP tasks on novels \cite{b1,b2,b3} has facilitated the analysis and downstream use of book artifacts \cite{b4,b5,b6,b7,b8}. These systems employ CoreNLP \cite{b9}, Stanza \cite{b10}, and SpaCy \cite{b11} for NLP annotation tasks. Further, methods to generate character embeddings and analyze relationships between characters \cite{b12,b13,b14,b15} provide additional information about characters apart from their mentions in the book.

\subsection{SSML Overview}
\label{sec:ssml_overview}
Speech Synthesis Markup Language (SSML) \cite{b17} is a markup language that provides a standard way to mark up text for the generation of synthetic speech, allowing for more fine-grained control over the generated audio. In this work, we are specifically interested in the \texttt{<prosody>} tag, which can be used to control three different attributes: pitch, volume, and rate of speech. We explore the role of audio in providing additional context and emotional cues \cite{b18}. An example of SSML is shown in Figure \ref{fig:sample_ssml}.

\begin{figure}[t]
    \centering
    \includegraphics[width=\linewidth]{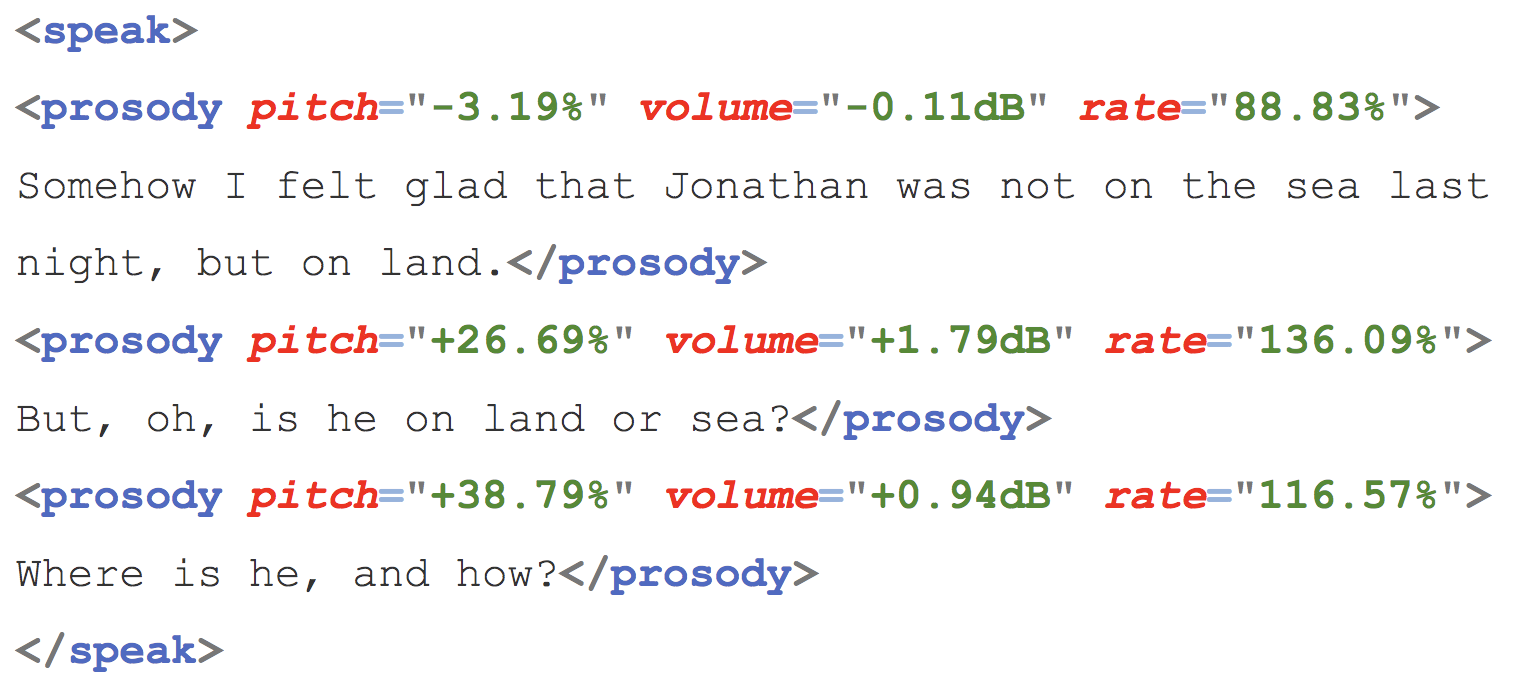}
    \caption{Sample SSML text with custom prosody attributes.}
    \label{fig:sample_ssml}
\end{figure}

\section{Audio/Text Alignment Methods}
\label{sec:audio-text-alignment}

We present a dataset of pairs of books and their corresponding audiobooks from the Project Gutenberg dataset, along with the extracted prosody data from the human-read audiobooks.
The dataset contains a total of 1806 chapters across 93 books.
We perform the following data processing steps to generate aligned data and extract the audio attributes.

\begin{itemize}
\item {\em Transcription} ---
We first split the audiobook into fragments using silence for segmentation. For each segment, we attempt to generate a transcript using Google Speech Recognition \footnote{\url{https://pypi.org/project/SpeechRecognition/}}. Note that this transcript is noisy, and we only use it to align with the appropriate chapter.

\item {\em Alignment} ---
We use the Gentle forced aligner\footnote{\url{https://github.com/lowerquality/gentle}} to generate alignments between  text and audio, including timestamp and phoneme information at word-level granularity.

\item {\em Audio Attribute Extraction} ---
\label{sec:audio_attrib}
To extract the pitch (fundamental frequency in Hz) and the volume (intensity in dB), we use the Python library Parselmouth\footnote{\url{https://parselmouth.readthedocs.io/en/stable/}}, with the timestep set to 0.01 second (compatible with the granularity of the alignment output from Gentle). To extract the rate of speech, we first compute the z-scores of the durations of all occurrences for each phoneme. Then we compute the mean of the phoneme duration z-scores for each audio segment (phrase / sentence), and the z-score of the resultant means.

\item {\em Sentence Segmentation} ---
\label{sec:phrase_chunk}
For each sentence, we generate a constituency parse tree using the Berkeley Neural Parser \cite{b17} with SpaCy \cite{b11}. We then split the sentence into phrases, using the constituent sentence (S) components and punctuation.

\end{itemize}

\section{Audiobook Reader Behavioral Analysis}
\label{sec:reader-behavior}

\begin{figure}[t]
\centering
\begin{subfigure}{\linewidth}
  \centering
  \includegraphics[width=0.9\linewidth]{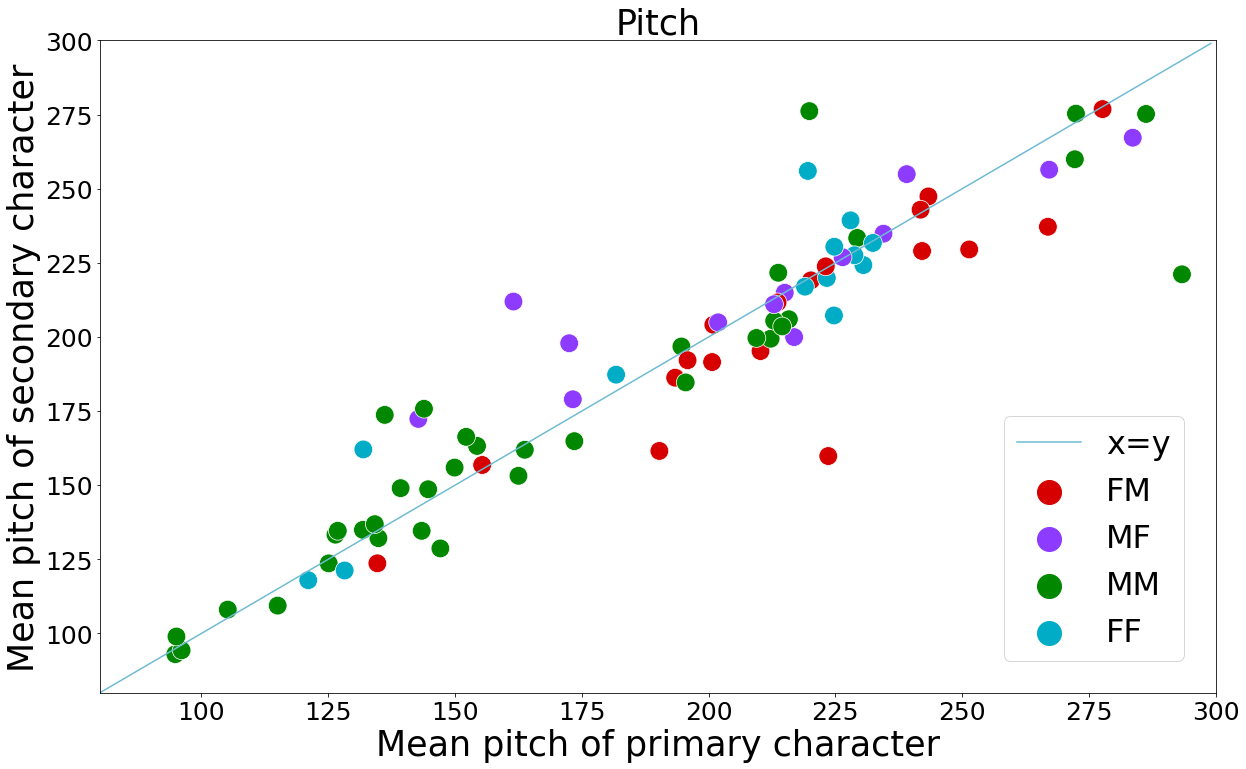}
  \caption{Pitch}
  \label{fig:pitch_main_character}
\end{subfigure}%
\\
\begin{subfigure}{\linewidth}
  \centering
  \includegraphics[width=0.9\linewidth]{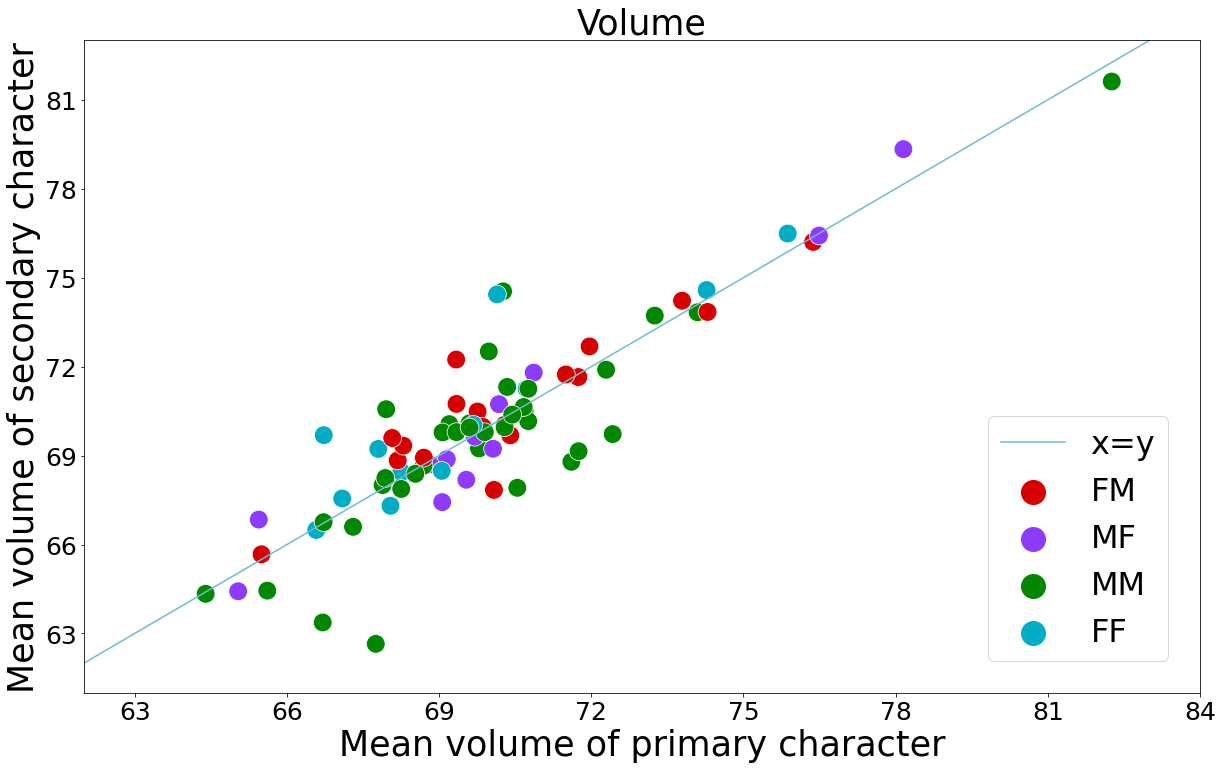}
  \caption{Volume}
  \label{fig:volume_main_character}
\end{subfigure}
\label{fig:pitch_and_volume_main_character}
\caption{Pitch and volume breakdown for the primary and secondary characters in each book, colored by gender pairs.}
\label{fig:pitch_and_volume_main_character}
\end{figure}

We compare pitch and volume for the most frequent and the second most frequent characters in each book. Figure \ref{fig:pitch_and_volume_main_character} shows pitch and volume of character pairs colored by gender. When the two main characters are of opposite gender, female character has a higher pitch and a lower volume compared to male character as most red dots are below the regression line, and most purple dots are above on both plots. 

We also compare the pitch and volume for dialogue utterances averaged separately across all the male and female characters in a book. For this comparison, we restrict our analysis to 62 books in which male and female characters each have at least 100 combined word utterances. For 52 out of 62 books, the mean pitch is higher for female character dialogue as compared to male (binomial p$=$3e-8). For 32 out of 62 books, the mean intensity (volume) is higher for male character dialogue as compared to female (binomial p$=$0.449).

\section{Predicting Prosody Attributes}
\label{sec:prosody-prediction}

We now address the task of predicting prosody attributes (pitch, volume, and rate) for each sentence or phrase, given a chapter text as input. We use Mean Squared Error (MSE) between human readers' actual prosody values and our predicted prosody values as accuracy measurement.

\subsection{Text Input Representations}
We experiment with three input embeddings: TF-IDF representations, Mean-pooled GloVe embeddings computed from the CommonCrawl pre-trained GloVe model (840B tokens), and MPNet embeddings \cite{b19} (\texttt{all-mpnet-base-v2}). We compute a single value for each prosody attribute per text segment, and compute the z-scores across all segments in the chapter.

We use a train-test split of 75-25\% across books, i.e. we use 69 books (1,392 chapters) in the training set and 24 books (414 chapters) in the test set. All the models are trained on a 2GHz CPU.

\subsection{Joint Prediction}
\label{sec:models}
Here we present the test metrics for various models at the phrase level, where we train a single common model to predict all three prosody attributes.

Using only a single phrase embedding as input (with no contextual information about the neighboring phrases), we predict the three attributes. Table \ref{tab:joint_phrase_non_contextual} shows the test MSE for the three attributes.
\begin{itemize}
    \item \textbf{LinReg:} We use linear regression as the baseline model, with the default parameters specified in sklearn\footnote{\url{https://scikit-learn.org/stable/modules/generated/sklearn.linear_model.LinearRegression.html}}.
    \item \textbf{MLP:} We use the MultiLinear Perceptron with the default parameters specified in sklearn\footnote{\url{https://scikit-learn.org/stable/modules/generated/sklearn.neural_network.MLPRegressor.html}}, with hidden layer sizes of (5,5), (10,10) and (20,20) respectively.
\end{itemize}

\begin{table}[htbp]
\small
\begin{center}
\setlength{\tabcolsep}{5pt}
\begin{tabular}{|r|r|rrr|}
\hline
\textbf{Embedding}      & \textbf{Model} & \textbf{Pitch}  & \textbf{Volume} & \textbf{Rate}   \\ \hline
\multirow{4}{*}{TF-IDF} & LinReg         & 0.9217          & 0.8789          & 0.8439          \\
                        & MLP (5, 5)     & 0.9103          & 0.8661          & 0.8264          \\
                        & MLP (10, 10)   & 0.9080          & 0.8609          & 0.8106          \\
                        & MLP (20, 20)   & 0.9101          & 0.8618          & 0.8067          \\ \hline
\multirow{4}{*}{GloVe}  & LinReg         & 0.9197          & 0.8644          & 0.9225          \\
                        & MLP (5, 5)     & 0.9173          & 0.8604          & 0.8905          \\
                        & MLP (10, 10)   & 0.9120          & 0.8488          & 0.8695          \\
                        & MLP (20, 20)   & 0.9014          & 0.8474          & 0.8615          \\ \hline
\multirow{4}{*}{MPNet}  & LinReg         & 0.8803          & 0.8048          & 0.8122          \\
                        & MLP (5, 5)     & 1.0001          & 1.0000          & 1.0000          \\
                        & MLP (10, 10)   & \textbf{0.8667} & 0.7817          & 0.7733          \\
                        & MLP (20, 20)   & 0.8668          & \textbf{0.7798} & \textbf{0.7666} \\ \hline
\end{tabular}
\end{center}
\centering
\caption{Test MSE for non-contextual joint prediction of prosody attributes}
\label{tab:joint_phrase_non_contextual}
\end{table}

\subsection{Sequential Prediction}
We use an LSTM model with input sequence length 2 or 3, with the following architecture: a Bidirectional LSTM layer (size 40), tanh-activation Dense layer (size 20), linear-activation (size 3). We use Mean Squared Error (MSE) as the loss function. We use a validation split of 15\%, train for 30 epochs, and select the model with the least validation loss.

\begin{table}[htbp]
\small
\setlength{\tabcolsep}{5pt}
\begin{center}
\begin{tabular}{|r|r|rrr|}
\hline
\textbf{Embedding} & \textbf{Model} & \textbf{Pitch}  & \textbf{Volume} & \textbf{Rate}   \\ \hline
GloVe              & LSTM(len 2)      & 0.8897          & 0.8260          & 0.8358          \\ \hline
MPNet              & LSTM(len 2)       & 0.8387          & 0.7540          & 0.7488          \\
                   & LSTM(len 3)       & \textbf{0.8362} & \textbf{0.7518} & \textbf{0.7449} \\ \hline
\end{tabular}
\end{center}
\caption{Test MSE for contextual joint prediction of prosody attributes with LSTM}
\label{tab:joint_phrase_lstm}
\end{table}
Table \ref{tab:joint_phrase_lstm} shows the test MSE for the three attributes. The LSTM model trained on MPNet phrase embeddings, with sequence length 3 shows the best performance. We use this model further downstream for text-to-speech audio generation and human evaluation.

\subsection{Prediction on Character Dialogue}
We observe that human readers read character dialogues more expressively than descriptive text.
\begin{figure}[htbp]
    \centering
    \includegraphics[width=\linewidth]{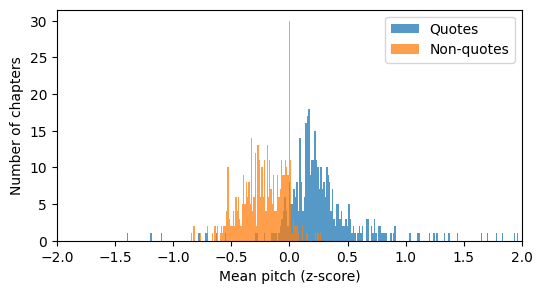}
    \caption{Mean ground-truth pitch z-score for quote and non-quote phrases from chapters in the test set}
    \label{fig:test_gt_pitch}
\end{figure}

Figure \ref{fig:test_gt_pitch} shows the distribution of the mean pitch z-scores for phrases that contain quotes or are a part of quotes, and phrases that do not contain quotes. Readers tend to use a higher pitch when reading dialogue and a lower pitch when reading descriptive text, as shown by the right and left shifts from 0 in the distribution.
\begin{figure}[htbp]
    \centering
    \includegraphics[width=\linewidth]{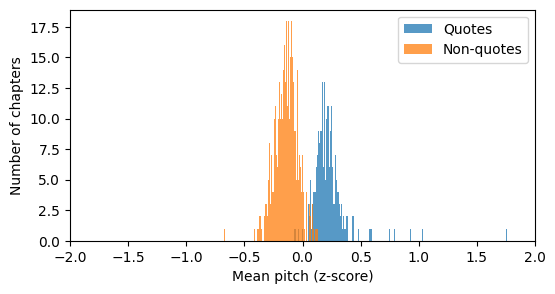}
    \caption{Mean predicted pitch z-score for quote and non-quote phrases from chapters in the test set}
    \label{fig:test_pred_pitch}
\end{figure}
Figure \ref{fig:test_pred_pitch} shows the same distributions for our model predictions. These follow a similar pattern, indicating that the model captures information about dialogue, even though we do not explicitly use this information during training.

We further examine this effect by looking at the prosody attribute prediction on dialogues only. Table \ref{tab:joint_phrase_lstm_dialogue} shows the MSE on a subset of the test data containing only dialogues. Predicting pitch and volume attributes on dialogues has a lower error compared to all input.

\begin{table}[htbp]
\small
\setlength{\tabcolsep}{5pt}
\begin{center}
\begin{tabular}{|r|r|rrr|}
\hline
\textbf{Embedding} & \textbf{Model} & \textbf{Pitch}  & \textbf{Volume} & \textbf{Rate}   \\ \hline
GloVe              & LSTM(len 2)      &  0.8568         &  0.7683      &   \textbf{0.7773}        \\ \hline
MPNet              & LSTM(len 2)       &  0.8073      &  0.6881     & 0.8291        \\
                   & LSTM(len 3)       & \textbf{0.8018}   &  \textbf{0.6872} & 0.8271 \\ \hline
\end{tabular}
\end{center}
\caption{Test MSE for contextual joint prediction of prosody attributes on dialogues}
\label{tab:joint_phrase_lstm_dialogue}
\end{table}

\subsection{Character Embeddings}
To include more subtle information about the nature of individual characters (e.g. hero vs.\ villain, old vs.\ young) that may inform actors' readings, we appended 908-dimensional character embeddings \cite{b12} to the input text representation and applied PCA for dimension reduction. Table \ref{tab:joint_phrase_lstm_dialogue_character_embed} shows the MSE on plain audio generated without SSML attributes (baseline) and enhanced audio generated with SSML attributes predicted from text and character embeddings. Appending character embeddings to the input results in improvements over the baseline, but not over our best models.

\begin{table}[htbp]
\small
\setlength{\tabcolsep}{3pt}
\begin{center}
\begin{tabular}{|r|r|rrr|}
\hline
\textbf{Embedding (Size)} & \textbf{Model} & \textbf{Pitch}  & \textbf{Volume} & \textbf{Rate}   \\ \hline
Baseline          &   N/A    &    1.3543      &   1.9550     &    2.3218       \\ \hline
GloVe +  CE (200) & LSTM(len 3)  &  1.0573  &  0.9559  &  1.2786   \\ \hline
GloVe + CE (100) & LSTM(len 3)  &   0.9181 &  0.8021  &  1.3043    \\ \hline
GloVe + CE (50) & LSTM(len 3)  &   0.9163   &  0.8169   &  1.4364        \\ \hline
MPNet + CE (200) & LSTM(len 3)  &  0.9384  & 0.8368  &  1.2711  \\ \hline
MPNet + CE (100) &  LSTM(len 3) &   0.8736 &  {\bf 0.7631} &  {\bf 1.0610}   \\ \hline
MPNet + CE (50) & LSTM(len 3) & {\bf 0.8734}  & 0.7709  &  1.0860     \\ \hline
\end{tabular}
\end{center}
\caption{Test MSE for joint prediction of prosody attributes on dialogues using character embeddings (CE) different PCA dimensions.}
\label{tab:joint_phrase_lstm_dialogue_character_embed}
\end{table}

\section{Generating and Evaluating Text-to-Speech Audiobooks}
\label{sec:human-evaluation}

We use the best performing model (LSTM with sequence length 3, and MPNet embeddings as input) to generate predictions for each book in the test set. We use a sliding window for inference, and compute the final prediction for each text input as the mean of its prediction in all windows.
To convert the predicted z-scores into SSML attributes, we compute the mean and standard deviation for the pitch and volume in each human-read audio file in the test set, and use these values to convert the prosody attributes into absolute values (Hz for pitch and dB for volume). Using the mean pitch/volume as reference, we convert these into relative values as required by the SSML specification (as described in Section \ref{sec:ssml_overview}). For rate of speech, we use a fixed value for mean and standard deviation (100 and 50 respectively) and perform the same computation.

We perform a human evaluation using Amazon Mechanical Turk. For each of the 24 books in the test set, we identify the phrases with the highest absolute values of predictions across the three prosody attributes. We then generate audio for this phrase text along with one context phrase before and after, using the Google Cloud Text-to-Speech API \footnote{\url{https://cloud.google.com/text-to-speech}} with the voice \texttt{en-US-Standard-A}. We produce two different audio variants for each text:
\begin{itemize}
    \item \textbf{Plain:} We enclose the text in a \texttt{<speak>} tag, without additional prosody information.
    \item \textbf{SSML-enhanced:} We enclose each phrase in a \texttt{<prosody>} tag with the three predicted attributes specified: see example in Figure \ref{fig:sample_ssml}.
\end{itemize}

For each pair of audios, we ask five annotators to rate which audio sounds better and more natural with intonations similar to human readers. Figure \ref{fig:bubble_plot} shows the results of the evaluation. For 12 out of 24 test sample pairs, a majority of the annotators preferred the SSML-enhanced version. 

\begin{figure}[htbp]
    \centering
    \includegraphics[scale=0.55]{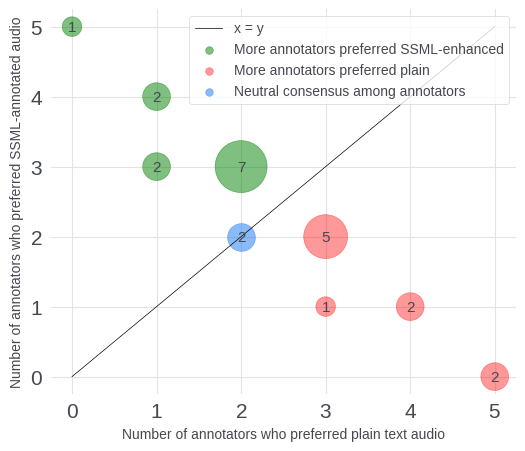}
    \caption{Distribution of books over annotator preference for  {\color[HTML]{036400} SSML} or {\color[HTML]{CB0000} Plain text}.}
    \label{fig:bubble_plot}
\end{figure}

\section{Conclusion}
\label{sec:conclusions}

We have developed an aligned text-to-audio book corpus, and deployed it to study both reader behavior and improve prosody models for TTS generation of audiobooks.
We have shown that actors customize voices to represent specific characters, and adopt more expressive voices than those generated by conventional TTS systems, motivating the development of full-book analysis methods to understand/predict a reader's goals in narrating a text.


\end{document}